\documentclass[aps,prfluids,reprint,superscriptaddress,nofootinbib]{revtex4-2}

\usepackage{graphicx}
\usepackage{amsmath,amssymb}
\usepackage{longtable}
\usepackage[usenames,dvipsnames]{xcolor}

\providecommand{\degree}{\ensuremath{^\circ}}

\makeatletter
\let\write@bibliographystyle\relax
\let\bibliography\@gobble
\makeatother

\begin{document}

\title{A floor and a ceiling for the advancing contact angle}

\author{Yong Sung Park}
\affiliation{Department of Civil, Urban and Environmental Engineering,
Seoul National University, Seoul 08826, Republic of Korea}
\affiliation{Institute of Construction and Environmental Engineering,
Seoul National University, Seoul 08826, Republic of Korea}

\date{\today}

\begin{abstract}
Reported maxima of the advancing contact angle scatter from 87\degree{} to
147\degree{} in a single systematic study, and liquids on surfaces with
static angles as low as
5\degree{} reach dynamic maxima near 90\degree{}. We show that both
observations follow from the hinged motion of the free surface near a
moving contact line. The local Stokes solution for a wedge rotating about
its contact line has two singular angles of opposite character: at
90\degree{} the hinged motion generates no wall shear and the rotation is
free, and at $\theta_h = 128.73\degree$, where $\tan 2\alpha = 2\alpha$,
a resonance with the $r^2$ eigensolution arrests the rotation through an
$r^2 \ln r$ term. Between the two angles lies a band that a transient
advancing angle cannot leave while the interface near the contact line
remains a quasi-steady wedge. A compilation of 68 liquid--solid systems
from nine sources and five configurations confirms the band and its two
exits: systems not limited by the apparatus, with $\theta_s \le 40\degree$, reach
87--119\degree{} regardless of chemistry, and every exceedance of
$\theta_h$ occurs either where chemistry places the static angle above the
band or where the flow leaves the quasi-steady regime. The record of a
waterline on a vertical wall exhibits the band within a single unsteady
experiment, together with a distinct quieting of the local contact angle
at the crossing of 90\degree{}.
\end{abstract}

\maketitle

\section{Introduction}
\label{sec:intro}

The largest angle that an advancing liquid front presents to a solid is
among the oldest measured quantities in wetting, and its published record
is scattered. Quetzeri-Santiago \emph{et al.}~\cite{QS2019} report maxima
from 87\degree{} to 147\degree{} across liquids, substrates, and
instruments, and the record as a whole scatters more widely still. The
scatter hides a regularity at its lower edge. Every liquid--substrate pair
they measured reaches a maximum advancing angle of at least 87\degree{},
including ethanol on glass with a static angle of $5\degree \pm 4\degree$;
in the compilation below, water on clean glass, with $\theta_s = 0$,
advances at 93\degree{}. A static angle varied over tens of degrees leaves
the dynamic maximum in place. Surface chemistry does not produce this
floor.

This paper locates a mechanism in the geometry of the motion itself.
A contact line passing from one steady state to another carries the
free surface with it, and near the contact line the transition is a
rotation of the interface about the line --- a hinged motion. The local
Stokes solution for this motion has two singular angles in the range of
interest, and they are singular in opposite senses. At $\alpha =
90\degree$ the hinged motion generates no shear stress on the wall, no
finite angular velocity balances a change of wall speed, and the rotation
is free: an angle in transition is swept through 90\degree{} and cannot
rest below it. At $\alpha = \theta_h = 128.73\degree$, the root of
$\tan 2\alpha = 2\alpha$, the forcing resonates with an eigensolution of
the wedge, the local solution acquires an $r^2 \ln r$ term, and the
torque required to rotate the interface through the angle diverges
logarithmically: the rotation is arrested. Gelderblom, Bloemen and
Snoeijer~\cite{GBS2012} obtained the same angle in the corner flow of an
evaporating droplet and named it $\theta_h$; we adopt their notation, and
we identify their angle with the upper edge of the measured record.
Between the two singularities lies a band, $90\degree \lesssim \alpha
\lesssim \theta_h$, and the claim of this paper is that a transient
advancing angle is confined to the band for as long as the interface near
the contact line remains a quasi-steady wedge. The confinement has two
exits, and the record uses both: chemistry may place the static angle
itself above the band, in which case the dynamics adds almost nothing, and
the flow may leave the quasi-steady regime, as it does in the first
instants of drop impact.

The claim operates at the scale of the apparent angle. The inner scales
of the moving contact line keep their established physics: the viscous
bending of the interface described by hydrodynamic
theory~\cite{Voinov1976,Cox1986}, the molecular kinetics of the
line~\cite{Blake2006}, and the entrainment of air at high
speed~\cite{Marchand2012} all act within the wedge whose opening angle the
hinge rotates. The band adds an outer, geometric constraint, and it is
distinguishable from the entrainment account by a single number:
entrainment thresholds respond to liquid viscosity with an exponent
measured at $1/3$ to $1/2$~\cite{Marchand2012}, and a geometric eigenangle
responds with an exponent of zero.

Section~\ref{sec:theory} derives the two angles from the similarity
solution. Section~\ref{sec:record} assembles 68 liquid--solid systems from
nine sources and five configurations, treats the apparatus limits in the
largest data blocks explicitly, and tests the band against the record, including a
regime test inside a single experiment. Section~\ref{sec:e1} examines an
unsteady case, the waterline of a solitary wave on a vertical wall, where
the hinged motion is the dominant physics. Section~\ref{sec:discussion}
states the predictions and the limits of the result.

\section{The hinged motion and its two singular angles}
\label{sec:theory}

We consider a two-dimensional wedge formed between the solid boundary and the
free surface near the moving contact line. At the onset of the transition
from a steady state with the speed at the solid boundary $U'$ and the contact
angle $\alpha$ to another with the speed $U' + \Delta U'$ and the contact angle
$\alpha + \Delta\alpha$, the free surface undergoes hinged motion with the
initial angular velocity $\omega'$. The flow field is described by the Stokes
equation with the Reynolds number based on the distance from the contact line
being arbitrarily small. We apply the Navier-slip boundary condition on the
solid boundary, which is convenient because the eigensolution
contributes to neither the bottom shear stress nor the pressure, and
zero-shear-stress and impermeability conditions on the free surface.

The Navier condition sets the slip velocity proportional to the shear stress and
so introduces the slip length $\ell = \mu'/\beta'$, where $\mu'$ is the dynamic
viscosity of the liquid and $\beta'$ the coefficient of sliding friction. We
scale lengths on $\ell$ and velocities on $U'$, so that
\begin{equation*}
  r = \frac{r'}{\ell} ,
  \qquad
  \omega = \frac{\ell\,\omega'}{U'} ,
  \qquad
  \Delta U = \frac{\Delta U'}{U'} .
\end{equation*}
With these variables the local similarity solution for the hinged motion
is~\cite{Moffatt1964,GBS2012}
\begin{equation}
  \Psi_{\omega s} = \omega r^2 \frac{g(\theta,\alpha)}{N(\alpha)},
  \qquad
  N(\alpha) = \frac{1}{\alpha}\left(2\alpha - \tan 2\alpha\right),
  \label{eq:solws}
\end{equation}
with $g(\theta,\alpha) = (2\alpha)^{-1}(\sin 2\theta - \tan 2\alpha \cos 2\theta
- 2\theta + \tan 2\alpha)$. Two angles in the range of interest make the
response to a hinged forcing singular. They do so in opposite senses.

\paragraph{The angle at which rotation is arrested.}
$N(\alpha)$ vanishes where $\tan 2\alpha = 2\alpha$. The first non-trivial root
is $2\alpha = 4.4934094579$, that is
\begin{equation}
  \alpha = \theta_h = 128.7267^\circ ,
  \label{eq:thetah}
\end{equation}
at which the exponent of the eigensolution is $\lambda_\omega = 2$ and the
eigensolution is resonant with the particular solution driven by the hinge. The
leading-order local solution then carries a factor $r^2\ln r$, so the torque
required to rotate the interface through this angle diverges logarithmically.
Gelderblom, Bloemen and Snoeijer~\cite{GBS2012} obtain the same angle, which
they write as $\theta_h = 128.7^\circ$, and describe the vanishing denominator
as signaling the breakdown of the local similarity solution. We adopt their
notation.

\paragraph{The angle at which rotation is free.}
Balancing the shear stress at the onset of the transition gives the angular
velocity of the hinged motion,
\begin{equation}
  \omega = \frac{\Delta U}{2}\,\frac{2\alpha - \tan 2\alpha}{\tan 2\alpha}.
  \label{eq:omega}
\end{equation}
As $\alpha \rightarrow 90^\circ$ ($=\pi/2$), $\tan 2\alpha \rightarrow 0$ and
$\omega$ diverges: the shear prefactor $\tan 2\alpha/(2\alpha - \tan 2\alpha)$
vanishes there, so the hinged motion generates no bottom shear stress and no
finite angular velocity can balance a finite $\Delta U$. We call this the
free-rotation singularity at $90^\circ$.

\begin{figure}[tb]
  \includegraphics[width=\columnwidth]{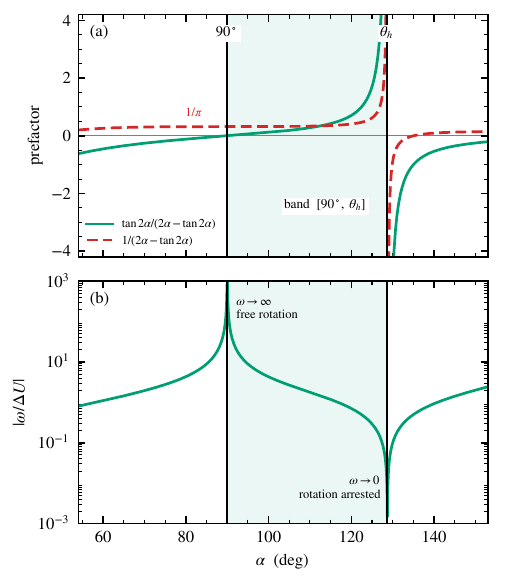}
  \caption{The two singular angles of the hinged motion. (a)~The
  $\alpha$-dependent prefactors of the local similarity solution. The solid
  curve is the shear prefactor $\tan 2\alpha/(2\alpha - \tan 2\alpha)$ and the
  dashed curve is the pressure prefactor $1/(2\alpha - \tan 2\alpha)$. Roughly
  over $72^\circ \leq \alpha \leq 108^\circ$ the former is close to a linear
  function of $\alpha - 90^\circ$ and the latter stays near $1/\pi$. Both
  diverge at $\theta_h$. (b)~The angular-velocity response~\eqref{eq:omega}.
  Between $90^\circ$ and $\theta_h$ it falls monotonically from divergence to
  zero, and the shaded band marks that interval. The script was written with
  Anthropic Claude (Opus~5) under the author's direction, and the plotted values
  were checked against the source data.}
  \label{fig:theory}
\end{figure}

Figure~\ref{fig:theory} places the two angles on one curve. The angular
velocity available to the hinge is unbounded at $90^\circ$ and vanishes at
$\theta_h$, and between them it decreases monotonically. A contact angle held
in transition therefore cannot rest below $90^\circ$, where the hinge is swept
through as fast as the forcing demands, and cannot be driven past $\theta_h$,
where the rotation is arrested. This is the band
\begin{equation}
  90^\circ \lesssim \alpha \lesssim \theta_h
  \label{eq:band}
\end{equation}
that the rest of this paper tests. Two properties of the bound matter for that
test. First, the arrest is carried by an $r^2\ln r$ term. The plateau therefore
moves with speed through a logarithm, and reported limiting angles should
scatter by several degrees. Second, the derivation assumes a quasi-steady wedge
with a flat free surface, so the band holds in that regime.
Section~\ref{sec:record} shows that the observations which leave the band are
the observations that leave the regime.

\section{The band against the record}
\label{sec:record}

\subsection{The compilation}

\begin{figure}[tb]
  \includegraphics[width=\columnwidth]{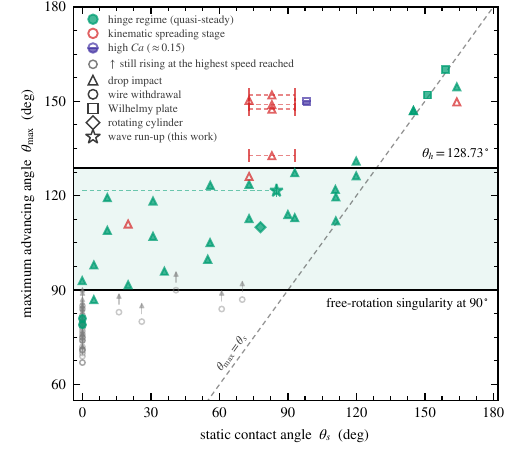}
  \caption{Maximum advancing angle against static contact angle, compiled across
  five configurations. Shading marks the band~\eqref{eq:band}. Filled symbols
  lie in the hinge regime. Open symbols were measured during the kinematic stage
  of drop spreading. The half-filled symbol was measured at
  $Ca \approx 0.15$. An upward arrow marks a value that was still rising at the
  highest speed the apparatus reached. A horizontal dashed bar carries the
  uncertainty in $\theta_s$. A bar closed at both ends spans a range consistent
  with a source that does not identify the substrate. A bar closed at one end
  only marks an upper bound on a single value, and its open end claims no lower
  limit. The star carries such a bar. It is the waterline of
  Sec.~\ref{sec:e1} and the one unsteady measurement in the compilation. Its
  colour is that of the hinge regime because the wedge near its contact line
  stays quasi-steady. The broken diagonal is $\theta_{\max} = \theta_s$.
  Sources are given in Table~\ref{tab:sources}. The script was written with
  Anthropic Claude (Opus~5) under the author's direction, and the plotted values
  were checked against the source data.}
  \label{fig:band}
\end{figure}

Figure~\ref{fig:band} collects $68$ liquid--solid systems from nine sources and
five configurations as the maximum advancing angle $\theta_{\max}$ against the
static contact angle $\theta_s$. The configurations are drop impact, wire
withdrawal, the Wilhelmy plate, a partially immersed rotating cylinder, and the
run-up of a solitary wave on a wall. Symbol shape denotes the configuration;
color and fill together denote the flow regime. Point-by-point sources and the
provenance of each coordinate are given in Table~\ref{tab:sources}.
The tables of Refs.~\cite{CR72441,CR72728} were transcribed by optical character
recognition, the published figures were digitized from their vector content, and
the record used in Sec.~\ref{sec:e1} was reprocessed, with scripts written under
the author's direction with Anthropic Claude (Opus~5); every transcribed row was
compared against a rendering of the original page, and the digitized values
against the numbers stated in the source texts.

Two kinds of marker carry bounds. An upward arrow marks a series whose contact
angle was still rising at the highest speed the apparatus reached, so its
$\theta_{\max}$ is a lower bound. A horizontal bar carries the uncertainty in
$\theta_s$, and its ends distinguish two cases. A bar closed at both ends spans
the range consistent with a source that does not identify the substrate. A bar
closed at one end and open at the other marks a system for which only an upper
bound on $\theta_s$ is reported: the value is a single one, and the bar does not
claim a lower end. The distinction governs how the figure may be read.

\subsection{Lower bounds in the wire-withdrawal data}

The wire-withdrawal measurements of Ellison and Tejada~\cite{CR72441} and of
Schwartz and Tejada~\cite{CR72728} are the largest single block in the
compilation, and in $28$ of their $30$ series the contact angle had not
saturated when the apparatus ran out of speed. Their reported angles are
therefore lower bounds. Read as maxima they would place $27$ points below
the floor. Read as bounds they place the same points on arrows pointing at it.
This block therefore serves as evidence neither for a limiting angle nor
against one.

\subsection{The floor}

Among the ten systems with $\theta_s \leq 40^\circ$ that were not limited by the
apparatus, $\theta_{\max}$
ranges from $87^\circ$ to $119^\circ$ while $\theta_s$ ranges from $0$ to
$36^\circ$; the gap $\theta_{\max} - \theta_s$ is between $82^\circ$ and
$108^\circ$. Water on clean glass, with $\theta_s = 0$, advances at
$93^\circ$. The extreme case reported by Quetzeri-Santiago \textit{et
al.}~\cite{QS2019} is ethanol on glass, with $\theta_s = 5^\circ \pm 4^\circ$
and $\theta_{\max} = 87^\circ \pm 4^\circ$. The static angle varies over this
whole range while the dynamic maximum stays at the floor.

The wire-withdrawal systems extend this observation to intermediate
wettability. Schwartz and Tejada report five systems with $\theta_{eq} > 0$;
their smooth-surface series have $\theta_{eq} = 16^\circ$, $26^\circ$,
$41^\circ$, $61^\circ$ and $70^\circ$, and reach $83^\circ$, $80^\circ$,
$90^\circ$, $84^\circ$ and $87^\circ$ respectively, every one of them still
rising when the apparatus ran out of speed. Across a $54^\circ$ span of static
angle these series approach the floor from below and stop there.

Two series in the Ellison and Tejada data turn over $1$--$2^\circ$ short of the others,
at $79^\circ$ and $81^\circ$, at the low speeds accessible to a wire of high
$\eta/\gamma$. The $\pm 1^\circ$ precision quoted for those measurements makes
them boundary cases. The floor is therefore a soft one. Low-viscosity systems
reach $87$ to $93^\circ$ and high-viscosity wire systems saturate a few degrees
below.

\subsection{The ceiling and the two ways out of it}

Thirteen systems in the compilation exceed $\theta_h$. Six of them were measured
in the hinge regime, and in five of those the static angle is itself above
$\theta_h$: the superhydrophobic surfaces of Kim \textit{et al.}~\cite{Kim2015}
at $\theta_s = 151^\circ$ and $159^\circ$, the non-wettable substrate of Bayer
and Megaridis~\cite{BM2006} at $\theta_E = 164^\circ$, and two liquids on the
Glaco-coated slides of Quetzeri-Santiago \textit{et al.} at
$\theta_s = 145^\circ$. These
points lie on the diagonal $\theta_{\max} = \theta_s$, and Kim \textit{et al.}
report that their advancing angles are insensitive to substrate velocity. Chemistry has placed the angle above the band and the
dynamics adds almost nothing.

One system in the hinge regime remains: water on PFAC$_8$, with
$\theta_s = 120^\circ$ and $\theta_{\max} = 131^\circ$. It exceeds $\theta_h$ by
$2.3^\circ$, against the $\pm 4^\circ$ uncertainty quoted by its own source.

The remaining seven exceedances were measured outside the hinge regime. Six of
them come from the kinematic stage of drop spreading and one from a Wilhelmy
plate at $Ca \approx 0.15$. In the first the free surface near the contact line
is the lamella edge in the instants after impact. In the second, viscous drag on
the plate distorts the apparent angle~\cite{PRF2019Wilhelmy}. Neither is the
quasi-steady flat wedge assumed in Sec.~\ref{sec:theory}.

\subsection{A regime test inside a single experiment}

A single source separates the two effects. Bayer and Megaridis divide their own spreading data into a
kinematic stage and a final stage, and they place the slow stage at
$V_{CL} < 0.2\,\mathrm{m\,s^{-1}}$. On their partially wettable substrate with
$\theta_E = 73^\circ$, the first spreading cycle at $We = 11.5$ reaches
$150.3^\circ$, well above $\theta_h$. The second spreading cycle of the same
drop on the same substrate, driven by the same impact, reaches $123.6^\circ$,
inside the band. The surface chemistry is held fixed throughout. The data
from the same experiment leave the band when the kinematic stage dominates and
return to it when the motion slows.

\subsection{A quantitative contrast with air entrainment}

The alternative account of a limiting advancing angle attributes it to the onset
of air entrainment. Its canonical measurement quantifies the material
dependence: Marchand \textit{et al.}~\cite{Marchand2012} report an
``unexpectedly weak dependence of entrainment speed on liquid viscosity'', with
the entrainment speed reduced by a factor $10$ while the viscosity is varied by
a factor $250$, which they fit with an exponent of $1/3$ or $1/2$. A geometric
eigenangle carries an exponent of zero. The two accounts therefore predict
exponents of $1/3$--$1/2$ and of $0$, and one set of measurements can
distinguish between them.

\subsection{Cautions on individual sources}

We read five entries in Table~\ref{tab:sources} differently from how they are
usually summarized. Each difference bears on the argument above.

(i) The statement of Bayer and Megaridis that the contact angle in the
inertia-controlled stage is constant ``regardless of the type of fluid'' traces
to Elliott and Ford~\cite{EF1972}, whose fluid-independent angle is the
$90^\circ$ that ends their first stage. Their maximum advancing angles,
$112^\circ$ and $120^\circ$, are described in the original as specific to a
given system and as sensitive to surfactant and to temperature.

(ii) The $112^\circ$ of Elliott and Ford lies within $1^\circ$ of the static
advancing angle of the same surface, $111^\circ$. A surface-chemical account of
that plateau is immediate. The point is included here for completeness.

(iii) Ellison and Tejada excluded glycerin and ethylene glycol from their
tabulation because the dynamic contact angles of those liquids ``exceeded
$90^\circ$ even at the low wire velocity limit of the apparatus''. The exclusion
mixes a physical threshold with an instrument limit and cannot be cited as
evidence for either.

(iv) Schwartz and Tejada selected systems with $\theta_{eq} < 90^\circ$, and the
Friz plots on which they report a departure from linearity near $90^\circ$ use
$\log \tan \theta_d$, and the singularity there is a coordinate singularity.

(v) The angle near $90^\circ$ that appears in molecular-dynamics
studies~\cite{PRF2020MD} is an equilibrium angle used as a wettability
parameter, and the $90^\circ$ in Ref.~\cite{Gupta2024} is a bound on the range
over which the experiment was operated. Both are quantities of a different kind,
and this paper cites them as such.
Ref.~\cite{Gupta2024} also remarks that the earlier dynamic studies it surveys
sit at $\theta_d > 90^\circ$, and describes the acute range as largely
unexplored. The remark runs in the direction of the floor, and we decline to use
it as evidence: a survey of coverage does not separate physics from the ease of
imaging an obtuse meniscus or from the systems a community selects, and the
standard of this section runs in both directions.

\section{An unsteady case: a waterline on a wall}
\label{sec:e1}

The data above are assembled from steady or quasi-steady measurements.
The hinged motion was introduced to describe a transition between steady states,
so the sharpest test of the band is an unsteady one. We use the
contact-line record of a solitary wave running up a vertical glass wall reported
in Ref.~\cite{PLC2012}, in which the equilibrium contact angle before each
experiment was below $85^\circ$ and the contact-line speed did not exceed
$0.04\,\mathrm{m\,s^{-1}}$, so that $Ca \lesssim 6\times 10^{-4}$ throughout.

\begin{figure}[tb]
  \includegraphics[width=\columnwidth]{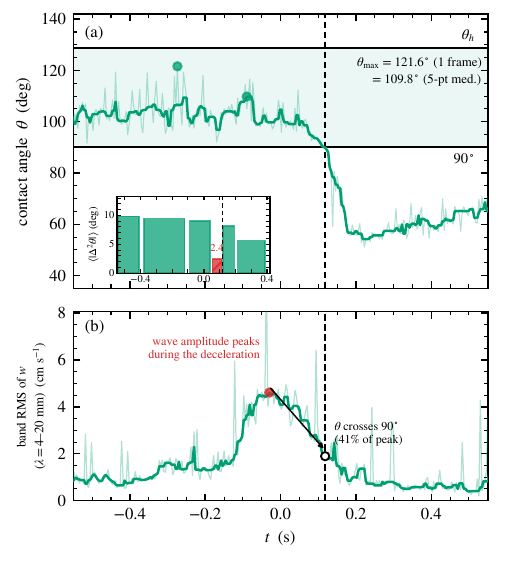}
  \caption{A waterline advancing and receding on a vertical wall. Time is
  measured from the instant at which the free-stream vertical velocity crosses
  zero, which is the origin used in Ref.~\cite{PLC2012}. The contact line
  reaches its highest point at $t = +0.110\,\mathrm{s}$. That reference plots
  $t/\sqrt{h_0/g}$, and $\sqrt{h_0/g} = 0.101\,\mathrm{s}$ for its water depth
  $h_0 = 10\,\mathrm{cm}$.
  (a)~Contact angle against time. The light trace is the record at
  $250\,\mathrm{Hz}$ and the heavy trace is its five-point median. Shading marks
  the band. The inset gives the frame-to-frame fluctuation of the local contact
  angle, $\langle|\Delta^2\theta|\rangle$, over successive intervals. The
  hatched bar is the interval immediately before the crossing of $90^\circ$,
  which the broken line marks. (b)~Root-mean-square amplitude of the free
  surface in the capillary band $\lambda = 4$--$20\,\mathrm{mm}$. The script was
  written with Anthropic Claude (Opus~5) under the author's direction, and the
  plotted values were checked against the source data.}
  \label{fig:e1}
\end{figure}

The angle stays inside the band for the whole of the advancing phase
[Fig.~\ref{fig:e1}(a)]. Its maximum is $121.6^\circ$ in a single frame of a
record whose frame-to-frame scatter is about $\pm 10^\circ$; the maximum of the
five-point median of the same record is $109.8^\circ$. By either estimate the
angle falls short of $\theta_h$ by $7^\circ$ to $19^\circ$ and never approaches
it. The waterline then descends, and the angle crosses $90^\circ$ at
$t = +0.118\,\mathrm{s}$, within two frames of the instant at which the contact
line reaches its highest point.

The crossing leaves a signature in the local contact angle. Over the
$67\,\mathrm{ms}$ before it the frame-to-frame fluctuation falls to
$2.41^\circ$, compared with $9.0$--$9.8^\circ$ over the preceding intervals, and
recovers to $8.22^\circ$ immediately after
[inset, Fig.~\ref{fig:e1}(a)]. The quieting cannot be an artifact of how the
angle is extracted. The sensitivity $\mathrm{d}\theta/\mathrm{d}s$ of the angle to the
measured surface slope is greatest exactly at $\theta = 90^\circ$, so noise
alone would produce a maximum of the fluctuation there. The observation is a
minimum.

\subsection{A capillary-wave interpretation, tested and rejected}

The obvious reading of that signature is that the crossing of $90^\circ$ radiates
capillary waves. We tested that reading against the same record, and it failed.
The amplitude of the free surface in the band
$\lambda = 4$--$20\,\mathrm{mm}$, which is where the gravity--capillary minimum
$c_{\min} = 0.231\,\mathrm{m\,s^{-1}}$ places waves excited on a
$20$--$50\,\mathrm{ms}$ timescale, is largest during the deceleration of the
contact line and has already fallen to $41\%$ of that maximum by the time the
angle crosses $90^\circ$ [Fig.~\ref{fig:e1}(b)]. No excursion after the crossing
comes near the deceleration peak, and the peak wavelength is $10$--$15\,$mm
throughout, so no new mode appears at the crossing. The radiated waves belong to
the deceleration of the contact line, as the original account of these
experiments states~\cite{PLC2012}.

The local signature survives. The fluctuation of the contact angle at the
contact line is a different quantity from the amplitude of waves propagating
away from it, and the two must be kept apart. Figure~\ref{fig:e1}(b) refutes the
radiated reading, and merging the two would turn it against the local one.

\section{Discussion}
\label{sec:discussion}

The band reads the existing record, and it asks something of future
measurements. Three predictions follow from the derivation. First, the
arrest at $\theta_h$ is carried by a logarithm, so the observed ceiling
drifts with speed and with the ratio of outer to inner scales by a few
degrees; reported plateaus should scatter accordingly, and they do ---
the floor is approached at 87--93\degree{} in low-viscosity systems and a
few degrees lower in high-viscosity wire systems, and the ceiling is
reported within a few degrees of $\theta_h$. Second, the edges of the
band carry a material exponent of zero. An experiment that holds the
geometry fixed and varies the viscosity by two orders of magnitude
separates the band from the entrainment account in a single plot: the
entrainment threshold moves with an exponent of $1/3$ to
$1/2$~\cite{Marchand2012}, and the band edges move only through the
logarithm. Third, departures from the band mark departures from the
quasi-steady wedge, and they end when quasi-steadiness returns; the two
spreading cycles of a single drop in Ref.~\cite{BM2006} show the exit and
the return within one experiment.

The dimensional form of Eq.~\eqref{eq:omega} separates the geometry from the
material. Restoring the scales gives
\begin{equation}
  \omega' = \frac{\Delta U'}{2\ell}\,
            \frac{2\alpha - \tan 2\alpha}{\tan 2\alpha} ,
  \label{eq:omegadim}
\end{equation}
with $\ell = \mu'/\beta'$ the slip length. The base speed $U'$ cancels, and the
hinge responds to the velocity increment alone. The liquid and the surface
enter only through $\ell$. It fixes the magnitude of $\omega'$, and it fixes
the argument $r'/\ell$ of the logarithm that carries the arrest. It is absent
from the angular factor, so the angles at which the response diverges and
vanishes are the same for every liquid and every surface. Those angles are the
edges of the band, and their independence of $\ell$ is what an exponent of zero
means.

The band also bounds a re-reading of past data. Measurements taken
during transients --- protocols that step the speed and read the angle
between steady states --- sample the hinge mode, and their plateaus near
the band edges record its two singular angles. The wire-withdrawal blocks of
Refs.~\cite{CR72441,CR72728} show a second, independent hazard: in 28 of
30 series the angle had not saturated when the apparatus ran out of
speed, and a lower bound cited as a maximum manufactures
evidence in either direction. The re-reading this paper proposes is
bounded by both observations: it concerns records near the band edges
taken during transients, and it asks first whether the apparatus, and
second whether the hinge, set the number.

The derivation carries its limits. The wedge is two-dimensional, the
solid boundary carries a Navier-slip condition, and the free surface is
flat and quasi-steady; a finite capillary number deforms the interface,
and the kinematic stage of impact replaces the wedge with a lamella edge.
The receding contact line is outside the scope of this paper, and its
singular structure is its own problem. The prefactors in
Eq.~\eqref{eq:omega} are those of the hinged wedge, and other unsteady
geometries will shift the approach to the band edges within the
logarithmic softness. What the two angles offer is a map: a floor that
chemistry cannot lower, a ceiling that dynamics alone cannot pass, and
marked borders whose crossings identify the physics that carried the angle
across --- chemistry above the band, unsteadiness outside the regime.

\section*{Data availability}
The compiled data behind Fig.~\ref{fig:band} and Table~\ref{tab:sources}, the
values digitized from the published figures of the cited sources, and the
scripts that produce Figs.~\ref{fig:theory}--\ref{fig:e1} and
Table~\ref{tab:sources} are openly available in Ref.~\cite{ZenodoDeposit}.
Figure~\ref{fig:theory} is computed from the closed-form expressions given in
Sec.~\ref{sec:theory} and uses no data. The contact-line record used in
Fig.~\ref{fig:e1} was obtained in the experiments reported in
Ref.~\cite{PLC2012} and is available from the author on reasonable request. The
transcription and digitization scripts read published papers and scanned
reports that are under third-party copyright, and those documents are not
redistributed.

\begin{acknowledgments}
This research was supported by the National Research Foundation of Korea (NRF)
grant funded by the Korea government (MSIT) [RS-2026-25475182]. The author also
acknowledges support from the Institute of Engineering Research at Seoul
National University.

The conception of this work---the identification of the hinged motion as the
mechanism, the two singular angles and the band between them, and the structure
of the argument---is the author's, who framed the draft. Anthropic Claude
(Fable~5 and Opus~5) drafted and revised the English prose of this paper, and
Claude Opus~4.5 and Google Gemini~3.1~Pro summarized the cited literature
through a batch interface, both under the author's direction; the author checked
every statement against the sources and is responsible for the text.
\end{acknowledgments}

\appendix
\section{Sources for Fig.~\ref{fig:band}}
\onecolumngrid
\scriptsize
\setlength{\LTcapwidth}{170mm}
\begin{longtable}{@{}l l c c l c@{\hspace{1.5mm}}l l c c l c@{}}
\caption{Sources for the points in Fig.~\ref{fig:band}, continued in the right-hand block. $\theta_s$ entries written $\leq x$ are upper bounds and $x$--$y$ are ranges consistent with the source; $\theta_{\max}$ entries written $\geq x$ were still rising at the highest speed the apparatus reached. Grade gives the provenance of ($\theta_s$, $\theta_{\max}$): T, a number stated in the text; F, read from a published figure. Abbreviations: aq.\ glycerol, aqueous glycerol of viscosity $4.7\,\mathrm{mPa\,s}$; 20 wt\% PEO, $63\,\mathrm{mPa\,s}$; SH, superhydrophobic; part., partially. Reg., flow regime: hinge, the hinge regime; kin., the kinematic spreading stage; $Ca$, the Wilhelmy plate at $Ca \approx 0.15$. Gr., grade.}\label{tab:sources}\\
\hline\hline
Ref. & System & $\theta_s$ & $\theta_{\max}$ & Reg. & Gr. & Ref. & System & $\theta_s$ & $\theta_{\max}$ & Reg. & Gr.\\
\hline\endfirsthead
\hline\hline
Ref. & System & $\theta_s$ & $\theta_{\max}$ & Reg. & Gr. & Ref. & System & $\theta_s$ & $\theta_{\max}$ & Reg. & Gr.\\
\hline\endhead
\hline\hline\endlastfoot
\cite{QS2019} & water / glass & 0 & 93 & hinge & FF & \cite{Yokoi2009} & distilled water / hydrophobic Si & 90 & 114 & hinge & TT\\
\cite{QS2019} & ethanol / glass / Glaco & 5 & 98 & hinge & FF & \cite{EF1972} & water / smooth paraffin wax & 111 & 112 & hinge & TT\\
\cite{QS2019} & ethanol / glass / Glaco & 5 & 87 & hinge & FF & \cite{Fell2011} & water, CTAB, CTAB+NaCl / sealant & 78 & 110 & hinge & TF\\
\cite{QS2019} & aq. glycerol / mica & 11 & 119.4 & hinge & FF & \cite{CR72728} & $\alpha$-bromonaphth. / nylon & 16 & $\geq 83$ & hinge & TT\\
\cite{QS2019} & water / mica & 11 & 109 & hinge & FF & \cite{CR72728} & n-octane / Teflon & 26 & $\geq 80$ & hinge & TT\\
\cite{QS2019} & aq. glycerol / cast acrylic & 31 & 118.3 & hinge & FF & \cite{CR72728} & methylene iodide / nylon & 41 & $\geq 90$ & hinge & TT\\
\cite{QS2019} & water / cast acrylic & 31 & 107.1 & hinge & FF & \cite{CR72728} & DEHS / Teflon & 61 & $\geq 84$ & hinge & TT\\
\cite{QS2019} & ethanol / cast acrylic & 36 & 96 & hinge & FF & \cite{CR72728} & water / nylon & 70 & $\geq 87$ & hinge & TT\\
\cite{QS2019} & ethanol / PMMA-like & 55 & 99.8 & hinge & FF & \cite{CR72441} & absolute alcohol / nylon & 0 & $\geq 79$ & hinge & TT\\
\cite{QS2019} & water / PMMA-like & 56 & 105.1 & hinge & FF & \cite{CR72441} & absolute alcohol / st. steel & 0 & $\geq 80$ & hinge & TT\\
\cite{QS2019} & aq. glycerol / PMMA-like & 56 & 123.3 & hinge & FF & \cite{CR72441} & absolute alcohol / titanium & 0 & $\geq 81$ & hinge & TT\\
\cite{QS2019} & aq. glycerol / Teflon & 93 & 127.3 & hinge & FF & \cite{CR72441} & hexadecane / PMMA & 0 & $\geq 76$ & hinge & TT\\
\cite{QS2019} & water / Teflon & 93 & 113 & hinge & FF & \cite{CR72441} & hexadecane / aluminum & 0 & 81 & hinge & TT\\
\cite{QS2019} & water / O$_2$-plasma Glaco & 111 & 122 & hinge & FF & \cite{CR72441} & hexadecane / nylon & 0 & $\geq 82$ & hinge & TT\\
\cite{QS2019} & aq. glycerol / O$_2$-plasma Glaco & 111 & 119.6 & hinge & FF & \cite{CR72441} & hexadecane / st. steel & 0 & $\geq 80$ & hinge & TT\\
\cite{QS2019} & aq. glycerol / PFAC$_6$/PFAC$_8$ & 120 & 126.3 & hinge & FF & \cite{CR72441} & hexadecane / titanium & 0 & $\geq 78$ & hinge & TT\\
\cite{QS2019} & water / PFAC$_6$/PFAC$_8$ & 120 & 131 & hinge & FF & \cite{CR72441} & freon TF / nylon & 0 & $\geq 79$ & hinge & TT\\
\cite{QS2019} & water / Glaco & 145 & 147 & hinge & TT & \cite{CR72441} & freon TF / st. steel & 0 & $\geq 75$ & hinge & TT\\
\cite{QS2019} & aq. glycerol / Glaco & 145 & 147 & hinge & TT & \cite{CR72441} & freon TF / titanium & 0 & $\geq 72$ & hinge & TT\\
\cite{BM2006} & water / wettable & 20 & 91.7 & hinge & TF & \cite{CR72441} & hexane / PMMA & 0 & $\geq 67$ & hinge & TT\\
\cite{BM2006} & water / wettable & 20 & 111 & kin. & TF & \cite{CR72441} & hexane / aluminum & 0 & $\geq 71$ & hinge & TT\\
\cite{BM2006} & water / part. wettable & 73--93 & 147.5 & kin. & FF & \cite{CR72441} & hexane / nylon & 0 & $\geq 70$ & hinge & TT\\
\cite{BM2006} & water / part. wettable & 73--93 & 148.9 & kin. & FF & \cite{CR72441} & hexane / st. steel & 0 & $\geq 67$ & hinge & TT\\
\cite{BM2006} & water / part. wettable & 73--93 & 152 & kin. & FF & \cite{CR72441} & iso-propanol / nylon & 0 & $\geq 69$ & hinge & TT\\
\cite{BM2006} & water / part. wettable & 73--93 & 132.7 & kin. & FF & \cite{CR72441} & iso-propanol / st. steel & 0 & $\geq 74$ & hinge & TT\\
\cite{BM2006} & water / part. wettable & 73 & 150.3 & kin. & TF & \cite{CR72441} & iso-propanol / titanium & 0 & $\geq 71$ & hinge & TT\\
\cite{BM2006} & water / part. wettable & 73 & 123.6 & hinge & TF & \cite{CR72441} & benzyl alcohol / PMMA & 0 & $\geq 74$ & hinge & TT\\
\cite{BM2006} & water / part. wettable & 73 & 126.1 & kin. & TF & \cite{CR72441} & benzyl alcohol / aluminum & 0 & $\geq 71$ & hinge & TT\\
\cite{BM2006} & water / part. wettable & 73 & 112.7 & hinge & TF & \cite{CR72441} & benzyl alcohol / nylon & 0 & $\geq 77$ & hinge & TT\\
\cite{BM2006} & water / non-wettable & 164 & 154.6 & hinge & TF & \cite{CR72441} & benzyl alcohol / st. steel & 0 & 79 & hinge & TT\\
\cite{BM2006} & water / non-wettable & 164 & 149.8 & kin. & TF & \cite{CR72441} & diethyl phthalate / aluminum & 0 & $\geq 75$ & hinge & TT\\
\cite{Kim2015} & PEO / smooth PTFE & 98 & 150 & $Ca$ & TT & \cite{CR72441} & diethyl phthalate / st. steel & 0 & $\geq 85$ & hinge & TT\\
\cite{Kim2015} & glycerin / SH PTFE & 151 & 152 & hinge & TF & \cite{CR72441} & diethyl phthalate / titanium & 0 & $\geq 84$ & hinge & TT\\
\cite{Kim2015} & PEO + glycerin / SH paint & 159 & 160 & hinge & TT & \cite{PLC2012} & tap water / glass wall & $\leq 85$ & 121.6 & hinge & TT\\
\end{longtable}
\normalsize
\twocolumngrid


\begin{thebibliography}{99}
\bibitem{QS2019} M.~A. Quetzeri-Santiago, A.~A. Castrej\'on-Pita, and
  J.~R. Castrej\'on-Pita, Phys. Rev. Lett. \textbf{122}, 228001 (2019).
\bibitem{GBS2012} H.~Gelderblom, O.~Bloemen, and J.~H. Snoeijer,
  J. Fluid Mech. \textbf{709}, 69 (2012).
\bibitem{Voinov1976} O.~V. Voinov, Fluid Dyn. \textbf{11}, 714 (1976).
\bibitem{Cox1986} R.~G. Cox, J. Fluid Mech. \textbf{168}, 169 (1986).
\bibitem{Blake2006} T.~D. Blake, J. Colloid Interface Sci. \textbf{299}, 1 (2006).
\bibitem{Marchand2012} A.~Marchand, T.~S. Chan, J.~H. Snoeijer, and B.~Andreotti,
  Phys. Rev. Lett. \textbf{108}, 204501 (2012).
\bibitem{Moffatt1964} H.~K. Moffatt, J. Fluid Mech. \textbf{18}, 1 (1964).
\bibitem{CR72441} A.~H. Ellison and S.~B. Tejada, NASA CR-72441 (1969).
\bibitem{CR72728} A.~M. Schwartz and S.~B. Tejada, NASA CR-72728 (1970).
\bibitem{Kim2015} J.~Kim, H.~P. Kavehpour, and J.~P. Rothstein,
  Phys. Fluids \textbf{27}, 032107 (2015).
\bibitem{BM2006} I.~S. Bayer and C.~M. Megaridis, J. Fluid Mech. \textbf{558},
  415 (2006).
\bibitem{PRF2019Wilhelmy} P.~Zhang and K.~Mohseni,
  Phys. Rev. Fluids \textbf{4}, 084004 (2019).
\bibitem{EF1972} T.~A. Elliott and D.~M. Ford, J. Chem. Soc., Faraday Trans. 1
  \textbf{68}, 1814 (1972).
\bibitem{PRF2020MD} J.-C. Fern\'andez-Toledano, T.~D. Blake, J.~De Coninck, and
  M.~Kandu\v{c}, Phys. Rev. Fluids \textbf{5}, 104004 (2020).
\bibitem{Gupta2024} C.~Gupta, A.~Choudhury, L.~D. Chandrala, and H.~N. Dixit,
  J. Fluid Mech. \textbf{1000}, A45 (2024).
\bibitem{PLC2012} Y.~S. Park, P.~L.-F. Liu, and I.-C. Chan,
  J. Fluid Mech. \textbf{707}, 307 (2012).
\bibitem{ZenodoDeposit} Y.~S. Park, \emph{Data and code for ``A floor and a
  ceiling for the advancing contact angle''}, Version~1.0,
  [Dataset and Software], Zenodo, 2026,
  \url{https://doi.org/10.5281/zenodo.21815877}.
\bibitem{Yokoi2009} K.~Yokoi, D.~Vadillo, J.~Hinch, and I.~Hutchings,
  Phys. Fluids \textbf{21}, 072102 (2009).
\bibitem{Fell2011} D.~Fell, G.~Auernhammer, E.~Bonaccurso, C.~Liu, R.~Sokuler,
  and H.-J. Butt, Langmuir \textbf{27}, 2112 (2011).
\end{thebibliography}
\end{document}